\documentclass[12pt]{article}
\usepackage{graphicx}
\usepackage{epsf,amsmath,bbold,amsfonts,stmaryrd}
\usepackage{mathrsfs}
\usepackage{appendix}
\usepackage{amssymb}
\usepackage{float}
\usepackage{color} 
\usepackage[colorlinks]{hyperref}
\hypersetup{pageanchor=false,citecolor=red,urlcolor=red}
\usepackage{indentfirst}
\usepackage[numbers,square,comma,compress,merge]{natbib}
\usepackage{multibib}
\usepackage[utf8]{inputenc}

\DeclareMathOperator*{\sumint}{%
\mathchoice%
  {\ooalign{$\displaystyle\sum$\cr\hidewidth$\displaystyle\int$\hidewidth\cr}}
  {\ooalign{\raisebox{.14\height}{\scalebox{.7}{$\textstyle\sum$}}\cr\hidewidth$\textstyle\int$\hidewidth\cr}}
  {\ooalign{\raisebox{.2\height}{\scalebox{.6}{$\scriptstyle\sum$}}\cr$\scriptstyle\int$\cr}}
  {\ooalign{\raisebox{.2\height}{\scalebox{.6}{$\scriptstyle\sum$}}\cr$\scriptstyle\int$\cr}}
}

\def\a{\alpha}

\def\d{\delta}
\def\D{\Delta}

\def\m{\mu}
\def\n{\nu}
\def\x{\xi}
\def\s{\sigma}

\def\op{\mathcal O}

\def\f{\frac}
\def\l{\left}

\def\p{\partial}
\def\r{\right}

\newcommand{\av}[1]{\langle #1\rangle}

\def\be{\begin{equation}}
\def\ee{\end{equation}}

\def\bea{\begin{eqnarray}}
\def\eea{\end{eqnarray}}

\def\ba{\begin{array}}
\def\ea{\end{array}}

\def\bc{\begin{center}}
\def\ec{\end{center}}

\begin{document}

\begin{titlepage}
\vspace{5cm}

\vspace{2cm}
\begin{center}

\bf \Large{CFT data and spontaneously broken conformal invariance}

\end{center}

\begin{center}
{\textsc {Georgios K. Karananas,$^{1, 2}$
    Mikhail Shaposhnikov\,$^1$}}
\end{center}

\begin{center}
{\it $^1$Institute of Physics\\
Laboratory of Particle Physics and Cosmology\\
\'Ecole Polytechnique F\'ed\'erale de Lausanne (EPFL)\\ 
CH-1015, Lausanne, Switzerland\\
\vspace{.2cm}
$^2$Arnold Sommerfeld Center\\
        Ludwig-Maximilians-Universit\"at M\"unchen\\
        Theresienstra{\ss}e 37, 80333 M\"unchen, Germany\\
}
\end{center}

\begin{center}
\texttt{\small georgios.karananas@physik.uni-muenchen.de} \\
\texttt{\small mikhail.shaposhnikov@epfl.ch} 
\end{center}

\vspace{2cm}

\begin{abstract}

We derive consistency conditions for the CFT data, which systems with
exact but spontaneously broken conformal invariance must satisfy.

\end{abstract}

\end{titlepage}

\leavevmode\\

\noindent\emph{\textbf{Introduction~--~}}The potential
phenomenological applications of scale (SI) and conformal invariance
(CI) in particle physics~\cite{Englert:1976ep,Bardeen:1995kv} and
cosmology~\cite{Wetterich:1987fm,Wetterich:1987fk,Wetterich:1994bg}
have been pointed out long ago. In the past years there has been a
resurgence of interest in this direction, see for
example~\cite{Meissner:2006zh,Shaposhnikov:2008xb,Shaposhnikov:2008xi,
Shaposhnikov:2009nk,Blas:2011ac,GarciaBellido:2011de,Bezrukov:2012hx,
Armillis:2013wya,Tavares:2013dga,Gretsch:2013ooa,Khoze:2013uia,Rubio:2014wta,
Boels:2015pta,Karam:2015jta,DiVecchia:2015jaq,Karananas:2016grc,Ferreira:2016vsc,Bianchi:2016viy,
Karananas:2016kyt,Karam:2016rsz,Ferreira:2016wem,Ferreira:2016kxi,Rubio:2017gty,
DiVecchia:2017uqn,Guerrieri:2017ujb,Tokareva:2017nng,Gillioz:2017ooc}.

Both of these symmetries apart from forbidding the presence of
dimensionful parameters in the action, also constrain heavily
the observables (correlation functions) of a
theory.\footnote{Obviously, conformal symmetry is more restrictive
than scale symmetry.} Usually, in unitary theories with
symmetry-preserving vacuum no distinction is made between SI and CI,
since the presence of the former implies the latter (in flat
spacetime)~\cite{Polchinski:1987dy,Luty:2012ww,Fortin:2012hn,Nakayama:2013is}.

Note that in interacting scale and conformal field theories (CFTs)
with symmetric vacuum there is no particle interpretation. As a
result, a crucial ingredient when it comes to utilizing SI or CI for
constructing theories that stand a chance of being phenomenologically
viable, is to require that they be spontaneously broken. Now, contrary
to what happens when the invariance under dilatations is linearly
realized, SI need not imply CI~\cite{Luty:2012ww}. In this case the
ground state is degenerate, as it contains a nontrivial flat direction
which is parametrized by the massless dilaton. This in turn forces the
cosmological constant to be zero and at the same time enables the
theory to accommodate massive excitations.

To the best of our knowledge, there has not been an attempt to study
generic theories exhibiting spontaneously broken SI or CI without a
known explicit Lagrangian formulation. In this paper, we will provide
a set of conditions that should be fulfilled by theories with exact,
but spontaneously broken CI. More specifically, we will derive
relations on the CFT data $=$ \{operator dimensions, Operator Product
Expansion (OPE) coefficients\} in the broken phase, which are
\emph{universal and independent of the specifics of a system}. It
should be noted that to study CFTs as a whole and not case by case, we
will not rely on a particular microscopic description; rather we will
be working solely with the OPE and correlators.

\newpage

\noindent\emph{\textbf{OPE and spontaneous breaking of conformal
symmetry~--~}}Let us start by assuming that the conformal symmetry is
spontaneously broken at a certain mass scale $v$---the vacuum
expectation value (vev) of the order parameter. As we already
mentioned, this is an important requirement if we wish for CI to be a
guiding principle for building realistic theories. We are going to
illustrate that by employing the OPE, it is possible to infer many
general properties of any unitary system that possesses a flat
direction along which the conformal symmetry is spontaneously
broken. More specifically, we will establish a set of consistency
conditions that need to be satisfied.

Our starting point is the OPE of two scalar primary operators, which
reads\,\footnote {It should be noted that, in principle, a theory
might contain more than one operator with the same scaling dimension.}
\be
\label{OPE_scal}
\op_i (x) \times \op_j(0)\sim \sum_k \frac{c_{ijk 
}}{|x|^{\D_{ijk}}}\op_k+\cdots \ .
\ee 
Here $c_{ijk}$ are the OPE coefficients, $|x|=\sqrt{x^\m
x_\m},~\op_l\equiv \op_l(0)$, $\D_{ijk}\equiv \D_i+\D_j-\D_k$,
with~$\D_l$ the dimensions, and the ellipses stand for
operators with nonzero spin, as well as descendants. For later
convenience, let us stress that no implicit summation over Latin
indices is assumed.

Consider a (unitary) four-dimensional CFT in which the conformal group
is spontaneously broken to its Poincar\'e subgroup, $SO(4,2)\to
ISO(3,1)$.\footnote{We will work exclusively in four-dimensional
Minkowski spacetime.}  This might happen, for instance, when some of
the (scalar) operators of the theory acquire a nonzero vev. To put it
differently, there exists a Poincar\'e-invariant ground
state, which we denote by $|0\rangle$, such
that
\be
\label{sym_brek}
\langle 0|  \op_i |0\rangle\equiv \av{\op_i}=\x_i \,v^{\D_i}\neq 0 \ ,
\ee
where $\x_i$'s are dimensionless parameters (and $v$ carries dimension
of mass).

From~\eqref{OPE_scal}, we find that when the OPE is sandwiched between
the symmetry-breaking vacuum $|0\rangle$, it yields
\be
\label{OPE_two-point}
\av{\op_i (x) \op_j(0)}\sim \sum_k \frac{c_{ijk
}}{|x|^{\D_{ijk}}}\av{\op_k}=\sum_k \frac{c_{ijk}}{|x|^{\D_{ijk}}}\x_k
v^{\D_k} \ .
\ee
Clearly, the only terms which survive and thus contribute to the
two-point function are the scalar operators. If the symmetry was not
broken, then only the unit operator ($\mathbb 1$) would be allowed to
acquire nonvanishing vev.  Since $\D_\mathbb{1}=0$, the expression
above would boil down to
\be
\av{\op_i (x) \op_j(0)}\sim\f{\d_{ij}}{|x|^{\D_i+\D_j}} \ ,
\ee
as it should.

Let us now insert a complete set of states in the left-hand side 
of~\eqref{OPE_two-point}, i.e.
\be
\av{\op_i (x) \op_j(0)}=\sumint_N ~\langle 0| \op_i(x) |
N\rangle\langle N |
\op_j(0)|0\rangle \ .
\ee
In the limit $x\to\infty$, due to the cluster decomposition
principle~\cite{Weinberg:1995mt,Weinberg:1996kr}, we will pick up only
the vacuum state. As a result, the above asymptotes to
\be
\label{CDP}
\lim_{x\to\infty} \av{\op_i (x) \op_j(0)} = \av{\op_i}\av{\op_j} =\x_i
\x_j\, v^{\D_i+\D_j} \ .
\ee
We can then conclude that the two-point
function~\eqref{OPE_two-point} formally yields
\be
\label{rel_1}
\x_i \x_j =\lim_{z\to 0}\sum_k c_{ijk}\, \x_k\, z^{\D_{ijk}} \ ,
\ee
where we introduced $z\equiv (v |x|)^{-1}$. This constitutes the first
relation that the CFT data must satisfy.  It is quite natural to
expect that the OPE will also contain operators whose scaling
dimension $\D_k$ is larger than $\D_i+\D_j$. In such case, and since
there is no \emph{a priori} reason for their corresponding OPE coefficients
to vanish, it is obvious that $\D_{ijk}<0$; consequently,
$z^{\D_{ijk}}$ will appear in the denominator of the consistency
condition~\eqref{rel_1}. The presence of such terms implies that the
infrared limit ($z\to 0$) should be taken only after the series have
been summed.  It should be stressed that whether this procedure is
mathematically well defined or not depends on the convergence of the
series in the above equation. Unfortunately, it is not known if this
is the case; the results of~\cite{Mack:1976pa,Pappadopulo:2012jk} (see
also~\cite{Luscher:1975js}), according to which the conformal OPE
indeed converges are not applicable here, for they were derived for
CFTs with unbroken vacuum. Nevertheless, in what follows we will
assume that the series in the right-hand side of~\eqref{rel_1} is
convergent at least in some finite domain of $z$, say at
$z\gtrsim\mathcal{O}(1)$, and that the result of summation can be
analytically continued to $z\to 0$.

Interestingly, the consistency relation~\eqref{rel_1} has been
presented previously by El-Showk and
Papadodimas~\cite{ElShowk:2011ag}, in the context of finite
temperature effects on CFTs.\footnote{We thank Jo\~ao Penedones for
bringing this to our attention.} Generally speaking, a rather natural
next step for our considerations would be to relax the requirement of
having a Lorentz-invariant vacuum. If the ground state preserves
spatial rotations only, this situation would be similar to what
happens in thermal CFTs, e.g. in high-temperature
QCD~\cite{Kharzeev:2007wb,Karsch:2007jc}. We leave this for future
work.

Coming back from this small digression, we note that, in principle, we
can go ahead and use the OPE to construct higher-order scalar
correlators. However, they will not provide us with further
information.\footnote{Three-point functions that involve two scalars
and, for example, the dilatational current $J_\m$, i.e.
\be
\av{O_i(x)O_j(0)J_\m(y)} \ ,
\ee
effectively reduce to the Ward identities and might give extra but
more complicated constraints, since these will involve double limits.}
To make this point more clear, let us consider for instance the
three-point function
\be
\av{\op_i(x)\op_j(0)\op_k(z)}\sim \sum_l  \frac{c_{ijl 
}}{|x|^{\D_{ijl}}}\av{\op_l(0)\op_k(z)}+\cdots \ ,
\ee
where, as before, we use the ellipses to denote terms involving the
derivatives of the operators etc. Following the previous logic, once
we insert complete sets of states between the three operators, it is
apparent that the above---upon using~\eqref{CDP}---boils down
to~\eqref{rel_1} in the deep infrared. It is not difficult to see that
this behavior persists in higher-order scalar functions.

Let us note in passing that the low-energy domain of the theory
contains \emph{only one Goldstone boson} $\pi$ associated with the
breaking of (SI and) CI, even though the number of broken generators
is five in total (one related to dilatations and four to special
conformal transformations). This fact does not depend on the details
of the symmetry-breaking mechanism and has been studied extensively in
the literature; the interested reader is referred
to~\cite{Salam:1970qk,Volkov:1973vd,Ogievetsky:1974,Ivanov:1975zq,Low:2001bw}
for further details. With this in mind, the next step is to consider
the implications of having $\pi$ in the spectrum. Assuming that the
theory contains no other massless scalar fields apart from the
dilaton, we expect that the vacuum amplitude of two operators will
contain a pole at vanishing
virtuality~\cite{Weinberg:1995mt,Weinberg:1996kr}.\footnote{We need
not require that the spin sectors of the system be gapped: modes with
nonzero spin cannot appear in~\eqref{gold_contr}, since their matrix
element with the state $\op_i|0\rangle$ vanishes
identically~\cite{Weinberg:1996kr}.} This becomes evident by inserting
into the scalar two-point function the ``resolution of unity,'' which
at the IR due to the domination of the dilaton can be approximated by
\be
\label{res_un}
1\underset{x\to\infty}{\sim}|0\rangle\langle0|+\int\f{d^3\vec
p}{2p_0(2\pi)^3}|\pi(p)\rangle\langle \pi(p)| \ .
\ee
A straightforward computation reveals that indeed
\be
\label{gold_contr}
 \av{\op_i (x) \op_j(0)}\underset{x\to\infty}{\sim} 
\av{\op_i}\av{\op_j}-\frac{\langle 0|\op_i |\pi\rangle\langle 
\pi|\op_j|0\rangle}{|x|^2} \ .
\ee 
Let us define the matrix element of the operator $\op_i$ between the
vacuum and the dilaton as\,\footnote{We use the conventional
(covariant) normalization for single-particle states
\be
\langle \pi(p) | \pi(p')\rangle=2p_0 (2\pi)^3 \d^{(3)}(\vec p-\vec
p\,') \ , 
\ee
with $p_0\equiv|\vec p|$. 
}
\be
\label{mat_el_vac_dil}
\langle 0| \op_i | \pi\rangle =f_i\,  v^{\D_i-1} \ ,
\ee
with $f_i$ a dimensionless coupling. Consequently, it is easy to see
that~\eqref{gold_contr}, upon using~\eqref{CDP}
and~\eqref{OPE_two-point}, leads to the second consistency condition
\be
\label{rel_2}
f_i f_j=\lim_{z\to0}\l[\f{1}{z^2}\l(\x_i \x_j-\sum_k c_{ijk}\, \x_k\,
z^{\D_{ijk}}\r)\r] \ .
\ee
It should be noted that Eq.~\eqref{rel_1} follows from Eq.~\eqref{rel_2}.
It is evident that the above includes an infinite number of new
parameters, $f_i$. As we will now show, these can in general be fixed
with the use of the Goldstone theorem. Let us note that even if a
conserved current $J_\m$ associated with dilatations
exists,\footnote{This assumption implies that the theory is local.}
this need not be invariant under translations; therefore, the
conventional proof (e.g.~\cite{Goldstone:1962es,Weinberg:1996kr})
might not be applicable here. The way out is to work directly with the
(improved) energy-momentum tensor
$T_{\m\n}$~\cite{Higashijima:1994}.\footnote{For an axiomatic approach
to the Goldstone theorem for symmetries whose currents are not
translationally invariant, see~\cite{Dothan1972,Ferrari:1973eq}.}

Lorentz invariance and $\p_\m T^{\m\n}=0$ dictate that the matrix
element of $T_{\m\n}$ between the vacuum
and the dilaton $\pi$, be of the following form,\footnote{Note that
a term proportional to $\eta_{\m\n}p^2$ is also
admissible in the matrix element~\eqref{rel_3_3}. However, this
contribution vanishes on shell.}
\be
\label{rel_3_3}
\l\langle 0| T_{\m\n}(0)  |\pi(p)\r \rangle=\f{1}{3}f_\pi v\, p_\m p_\n \ ,
\ee 
with $f_\pi$ the dimensionless dilaton decay constant and the factor
of $1/3$ was added for later convenience. It can then be shown that
the expectation value of the commutator between the energy-momentum
tensor and an operator reads
\be 
\label{rel_3_1} \l\langle [T_{\m\n},\op_i]\r\rangle=\f{i}{3}f_\pi
f_iv^{\D_i}\,\p_\m \p_\n G(x)\ ,
\ee
where $G(x)$ is the (massless) Green's function. Provided that
$J_\m\equiv x^\n T_{\m\n}$, then the charge associated with
dilatations can be written as a particular moment of the
energy-momentum tensor
\be
\label{rel_3_2}
D=\int d^3\vec x\, x^\m T_{0\m} \ ,
\ee
while, by definition, 
\be
\label{rel_3_3}
\av{[D,\op_i]}=i\x_i \D_i v^{\D_i} \ .
\ee
Consequently, from the relations~\eqref{rel_3_1}--\eqref{rel_3_3}, it
easily follows that
\be
\label{rel_3}
f_i=\frac{\x_i\, \D_i}{f_\pi} \ .
\ee
We observe that~\eqref{rel_1},~\eqref{rel_2} and~\eqref{rel_3},
constitute a system of equations for $\x_i$ and $\D_i$. If
a nontrivial solution exists, this can serve as an indication that
the CFT data describes a system that exhibits the symmetry breaking
pattern $SO(4,2)\to ISO(3,1)$.

At this point, we would like to turn our attention
to the energy-momentum tensor, an operator of particular importance as
far as CFTs are concerned. The relevant for our considerations terms
in the two-point correlator of $T_{\m\n}$ with itself are
\be
\label{OPE_EM}
\av{T_{\m\n}(x)T_{\lambda\s}(0)}=\displaystyle\sum_k 
\mathscr{T}_{\m\n\lambda\s}\f{\x_kv^{\D_k}}{|x|^{8-\D_k}} \ .
\ee
To keep the expression short, we introduced the most general
Lorentz-covariant structure consistent with the symmetries of the
energy-momentum tensor (see also~\cite{Osborn:1993cr,Dorigoni:2009ra})
\be
\label{OPE_EM_tensor}
\begin{aligned}
\mathscr{T}_{\m\n\lambda\s}&=\text{a}^T_{k}A_{\m\n\lambda\s}+
b^T_{k}B_{\m\n\lambda\s}\\
&+\f{1}{|x|^2}\l(c^T_{k}C_{\m\n\lambda\s}
+d^T_{k}D_{\m\n\lambda\s}\r)
+e^T_{k}\f{E_{\m\n\lambda\s}}{|x|^4} \ ,
\end{aligned}
\ee
where
\be
\label{OPE_EM_tensor_2}
\begin{aligned}
&A_{\m\n\lambda\s}=\eta_{\m\n}\eta_{\lambda\s},\\
&B_{\m\n\lambda\s}=\eta_{\m\lambda}\eta_{\n\s}+
\eta_{\m\s}\eta_{\n\lambda},\\
&C_{\m\n\lambda\s}=
\eta_{\m\n}x_\s x_\lambda+\eta_{\lambda\s}x_\n x_\m,\\
&D_{\m\n\lambda\s}=\eta_{\lambda\m}x_\n x_\s +\eta_{\m\s}
x_\n x_\lambda+\eta_{\n\s}x_\m x_\lambda +\eta_{\n\lambda}
x_\m x_\s,\\
&E_{\m\n\lambda\s}=x_\m x_\n x_\lambda x_\s,
\end{aligned}
\ee
and $\eta_{\m\n}$ is the Minkowski metric. From the vanishing of the
divergence
\be
\p^\m\av{T_{\m\n}(x)T_{\lambda\s}(0)}=0 \ ,
\ee
we obtain 
\bea
&&2d^T_k-(5-\D_k)c^T_k-(8-\D_k)\text{a}^T_k=0 \ ,\nonumber\\
&&c^T_k-(4-\D_k)d^T_k-(8-\D_k)b^T_k=0 \ ,\\
&&(10-\D_k)(c^T_k+2d^T_k)+(5-\D_k)e^T_k=0 \ ,\nonumber 
\eea
which must hold for each $k$.

In addition, since $T^\m_\m=0$, it immediately follows that for all
$k$'s
\be
16\text{a}^T_k+8b^T_k+8c^T_k+4d^T_k+e^T_k=0 \ ,
\ee
where we used~\eqref{OPE_EM}--\eqref{OPE_EM_tensor_2}. Note that from
the above algebraic equations we can express four of the coefficients
in terms of one, say $\text{a}_k^T$. Once this is effectuated, we find
that the two-point function~\eqref{OPE_EM} boils down to
\be
\label{OPE_EM_a}
\av{T_{\m\n}(x)T_{\lambda\s}(0)}=\displaystyle\sum_k 
\widetilde{\mathscr{T}}_{\m\n\lambda\s}\f{\tilde{\a}_k\,\text{a}_k^T\x_kv^{\D_k}}
{|x|^{8-\D_k}} \ ,
\ee
with $\tilde{\a}_k=(10-8\D_k+\D_k^2)^{-1}$, while
\be
\label{OPE_EM_tensor_a}
\begin{aligned}
\widetilde{\mathscr{T}}_{\m\n\lambda\s}&=\a_k^{(1)}A_{\m\n\lambda\s}-
\a_k^{(2)} B_{\m\n\lambda\s}\\
&+\f{1}{|x|^2}\l(\a_k^{(3)} C_{\m\n\lambda\s}
+\a_k^{(4)} D_{\m\n\lambda\s}\r)-2\a_k^{(5)}\f{E_{\m\n\lambda\s}}{|x|^4} \ ,
\end{aligned}
\ee
and we have defined
\be
\begin{aligned}
\label{constr_OPE_coeffs}
&\a_k^{(1)}= 1,~\a_k^{(2)}=20-12\D_k+\f{3}{2}\D_k^2,
~\a_k^{(3)}=(8-\D_k)\D_k,\\
&\a_k^{(4)}=40-17\D_k+\f{3}{2}\D_k^2,
~\a_k^{(5)}=80-18\D_k+\D_k^2 \ . 
\end{aligned}
\ee
As a sanity check, if the symmetry were linearly realized, then only
the unity would contribute. In such a case,
Eqs.~\eqref{OPE_EM_a}--\eqref{constr_OPE_coeffs} dictate that up to
irrelevant numerical factors
\be
\begin{aligned}
&\av{T_{\m\n}(x)T_{\lambda\s}(0)}_\text{unbroken}\\
&~~~\propto
\text{a}^T_\mathbb{1}\l(\f{1}{2}\l(I_{\m\lambda}I_{\n\s}+I_{\n\lambda}I_{\m\s}\r
)-\f{1}{4}\eta_{\m\n}\eta_{\lambda\s}\r)\f{1}{|x|^8} \ ,
\end{aligned}
\ee
with $I_{\m\n}\equiv\eta_{\m\n}-2x_\m x_\n/|x|^2$, and
$a_\mathbb{1}^T$ an overall coefficient setting the scale of the
two-point function.\footnote{Note that instead of the symbol
$\text{a}^T_\mathbb{1}$, $C_T$ is most commonly used in the
literature.}

If we now consider the low-energy limit, we end up
with the following constraints on $\text{a}_k^T$
\be
\label{constr_ten_ir}
\lim_{z\to 0}\sum_k \a_k^{(n)}\,\tilde{\a}_k\,\text{a}^T_k\x_k
z^{8-\D_k}=0 \ ,~~~n=1,\ldots,5 \ .
\ee

Yet another set of conditions relating the OPE coefficients of
$T_{\m\n}$ can be obtained by considering its interaction
with the dilaton. This practically amounts to
plugging~\eqref{res_un} into the left hand side of the
correlator~\eqref{OPE_EM_a}. A (long but) straightforward computation
gives 
\be
\label{constr_ten_dil}
\lim_{z\to 0}\sum_k \a_k^{(n)}\,\tilde{\a}_k\,\text{a}^T_k\x_k
z^{2-\D_k}=C^{(n)}f_\pi^2 \ ,~~~n=1,\ldots,5 \ ,
\ee
with
\be
C^{(1)}=-C^{(2)}=\f{8}{9} ,~C^{(3)}=C^{(4)}=-\f{16}{3}
,~C^{(5)}=-\f{64}{3} \ .
\ee
Like it happened before, Eq.~\eqref{constr_ten_ir} is a consequence
of~\eqref{constr_ten_dil}.

Before moving to the conclusions, let us for completeness discuss briefly what can be
deduced by considering the OPE of (translationally invariant) vector
operators. The OPE contains, among others, the following terms
\be
\label{OPE_vector}
V^i_\m(x)\times V^j_\n(0)\supset
\sum_k\l(\text{a}^V_{ijk}\eta_{\m\n}+b^V_{ijk}\f{x_\m x_\n}{|x|^2}\r)
\f{\op_k}{|x|^{\D_{ijk}}} \ .
\ee
By taking the average in the symmetry-breaking vacuum, we see that in the
IR ($x\to\infty$) the left-hand side must be zero due to Lorentz invariance. As a
consequence, we obtain a set of constraints that the OPE
coefficients $\text{a}^V_{ijk}$ and $b^V_{ijk}$---provided that they
are not trivial---should satisfy
\be
\label{constr_vector}
\lim_{z\to 0}\sum_k \text{a}^V_{ijk}\,\x_k \,z ^{\D_{ijk}}=0 \
,~~~\lim_{z\to 0}\sum_k b^V_{ijk}\,\x_k \,z ^{\D_{ijk}}=0 \ .
\ee 
As before, $z=(v|x|)^{-1}$. Note that for higher-spin operators, owing to
the fact that for $z\to 0$ (equivalently, $x\to\infty$) their vev's
must also vanish, a similar type of relations can be obtained; see for
example the ones we presented for the energy-momentum tensor.

\vspace{.2cm}

\noindent\emph{\textbf{Conclusions~--~}}In the present short paper, we
reported on constraints that the CFT data should satisfy when a system
possesses a flat direction, and thus, exhibits nonlinearly realized
conformal invariance. Our considerations are very general, for they
only require knowledge of the operator spectrum of a theory, their
corresponding anomalous dimensions and the OPE coefficients. What
remains to be seen is if the relations we presented can be used to
identify the allowed regions of the phase portrait of CFTs with
spontaneously broken symmetry; the main challenge is the convergence
of the OPE and whether it is possible to analytically continue it into
the infrared regime.

An ideal testing ground for our findings would be theories such as
$\mathcal N=4$ super Yang-Mills, which is known to possess exact but
spontaneously broken conformal invariance at the Coulomb branch. It
would be very interesting to confront our results with the ones for
the full spectrum of the anomalous dimensions for this
theory~\cite{Gromov:2009tv,Gromov:2009bc,Gromov:2009zb}.

Although a bit tangent to this paper, let us mention that on the
phenomenological side, spontaneously broken scale and conformal
symmetries have served as a guiding principle for constructing
realistic theories able to describe our Universe from its very early
stages up until the present day. In their context, the hierarchy and
cosmological constant problems can be viewed from a fresh
perspective. Their potential resolution might be achieved under
certain extra assumptions on the UV dynamics, such as the absence of
new particle thresholds between the electroweak and Planck
scales~\cite{Shaposhnikov:2007nj,Shaposhnikov:2008xi,Giudice:2013nak}
(see also~\cite{Karananas:2017mxm} for an implementation of this idea
in grand unified theories).

\vspace{.2cm}

\noindent\emph{\textbf{Acknowledgements~--~}}We thank Jo\~ao Penedones
and Slava Rychkov for helpful discussions and important comments on
the paper. This work was supported by the ERC-AdG-2015 grant
694896. The work of G.K.K. and M.S. was supported partially by the
Swiss National Science Foundation.

\vspace{.5cm}

\noindent\makebox[\linewidth]{\resizebox{.8\linewidth}{1.2pt}{$\blacklozenge$}}\
\bigskip {\small \vspace{-2cm}
\bibliographystyle{utphys}
\bibliography{Conformal_SSB_Correlators}{} }

\providecommand{\href}[2]{#2}\begingroup\raggedright\begin{thebibliography}{10}

\bibitem{Englert:1976ep}
F.~Englert, C.~Truffin, and R.~Gastmans, ``{Conformal Invariance in Quantum
  Gravity},''
\href{http://dx.doi.org/10.1016/0550-3213(76)90406-5}{{\em Nucl. Phys.}
  {\bfseries B117} (1976) 407--432}.

\bibitem{Bardeen:1995kv}
W.~A. Bardeen, ``{On naturalness in the standard model},'' in {\em {Ontake
  Summer Institute on Particle Physics Ontake Mountain, Japan, August
  27-September 2, 1995}}.
\newblock
\url{http://lss.fnal.gov/cgi-bin/find_paper.pl?conf-95-391}.
\newblock

\bibitem{Wetterich:1987fm}
C.~Wetterich, ``{Cosmology and the Fate of Dilatation Symmetry},''
\href{http://dx.doi.org/10.1016/0550-3213(88)90193-9}{{\em Nucl. Phys.}
  {\bfseries B302} (1988) 668--696}.

\bibitem{Wetterich:1987fk}
C.~Wetterich, ``{Cosmologies With Variable Newton's 'Constant'},''
\href{http://dx.doi.org/10.1016/0550-3213(88)90192-7}{{\em Nucl. Phys.}
  {\bfseries B302} (1988) 645--667}.

\bibitem{Wetterich:1994bg}
C.~Wetterich, ``{The Cosmon model for an asymptotically vanishing time
  dependent cosmological 'constant'},'' {\em Astron. Astrophys.} {\bfseries
  301} (1995) 321--328,
\href{http://arxiv.org/abs/hep-th/9408025}{{\ttfamily arXiv:hep-th/9408025
  [hep-th]}}.

\bibitem{Meissner:2006zh}
K.~A. Meissner and H.~Nicolai, ``{Conformal Symmetry and the Standard Model},''
  \href{http://dx.doi.org/10.1016/j.physletb.2007.03.023}{{\em Phys. Lett.}
  {\bfseries B648} (2007) 312--317},
\href{http://arxiv.org/abs/hep-th/0612165}{{\ttfamily arXiv:hep-th/0612165
  [hep-th]}}.

\bibitem{Shaposhnikov:2008xb}
M.~Shaposhnikov and D.~Zenhausern, ``{Scale invariance, unimodular gravity and
  dark energy},'' \href{http://dx.doi.org/10.1016/j.physletb.2008.11.054}{{\em
  Phys. Lett.} {\bfseries B671} (2009) 187--192},
\href{http://arxiv.org/abs/0809.3395}{{\ttfamily arXiv:0809.3395 [hep-th]}}.

\bibitem{Shaposhnikov:2008xi}
M.~Shaposhnikov and D.~Zenhausern, ``{Quantum scale invariance, cosmological
  constant and hierarchy problem},''
  \href{http://dx.doi.org/10.1016/j.physletb.2008.11.041}{{\em Phys. Lett.}
  {\bfseries B671} (2009) 162--166},
\href{http://arxiv.org/abs/0809.3406}{{\ttfamily arXiv:0809.3406 [hep-th]}}.

\bibitem{Shaposhnikov:2009nk}
M.~E. Shaposhnikov and F.~V. Tkachov, ``{Quantum scale-invariant models as
  effective field theories},''
\href{http://arxiv.org/abs/0905.4857}{{\ttfamily arXiv:0905.4857 [hep-th]}}.

\bibitem{Blas:2011ac}
D.~Blas, M.~Shaposhnikov, and D.~Zenhausern, ``{Scale-invariant alternatives to
  general relativity},''
  \href{http://dx.doi.org/10.1103/PhysRevD.84.044001}{{\em Phys. Rev.}
  {\bfseries D84} (2011) 044001},
\href{http://arxiv.org/abs/1104.1392}{{\ttfamily arXiv:1104.1392 [hep-th]}}.

\bibitem{GarciaBellido:2011de}
J.~Garcia-Bellido, J.~Rubio, M.~Shaposhnikov, and D.~Zenhausern,
  ``{Higgs-Dilaton Cosmology: From the Early to the Late Universe},''
  \href{http://dx.doi.org/10.1103/PhysRevD.84.123504}{{\em Phys. Rev.}
  {\bfseries D84} (2011) 123504},
\href{http://arxiv.org/abs/1107.2163}{{\ttfamily arXiv:1107.2163 [hep-ph]}}.

\bibitem{Bezrukov:2012hx}
F.~Bezrukov, G.~K. Karananas, J.~Rubio, and M.~Shaposhnikov, ``{Higgs-Dilaton
  Cosmology: an effective field theory approach},''
  \href{http://dx.doi.org/10.1103/PhysRevD.87.096001}{{\em Phys. Rev.}
  {\bfseries D87} no.~9, (2013) 096001},
\href{http://arxiv.org/abs/1212.4148}{{\ttfamily arXiv:1212.4148 [hep-ph]}}.

\bibitem{Armillis:2013wya}
R.~Armillis, A.~Monin, and M.~Shaposhnikov, ``{Spontaneously Broken Conformal
  Symmetry: Dealing with the Trace Anomaly},''
  \href{http://dx.doi.org/10.1007/JHEP10(2013)030}{{\em JHEP} {\bfseries 10}
  (2013) 030},
\href{http://arxiv.org/abs/1302.5619}{{\ttfamily arXiv:1302.5619 [hep-th]}}.

\bibitem{Tavares:2013dga}
G.~Marques~Tavares, M.~Schmaltz, and W.~Skiba, ``{Higgs mass naturalness and
  scale invariance in the UV},''
  \href{http://dx.doi.org/10.1103/PhysRevD.89.015009}{{\em Phys. Rev.}
  {\bfseries D89} no.~1, (2014) 015009},
\href{http://arxiv.org/abs/1308.0025}{{\ttfamily arXiv:1308.0025 [hep-ph]}}.

\bibitem{Gretsch:2013ooa}
F.~Gretsch and A.~Monin, ``{Perturbative conformal symmetry and dilaton},''
  \href{http://dx.doi.org/10.1103/PhysRevD.92.045036}{{\em Phys. Rev.}
  {\bfseries D92} no.~4, (2015) 045036},
\href{http://arxiv.org/abs/1308.3863}{{\ttfamily arXiv:1308.3863 [hep-th]}}.

\bibitem{Khoze:2013uia}
V.~V. Khoze, ``{Inflation and Dark Matter in the Higgs Portal of Classically
  Scale Invariant Standard Model},''
  \href{http://dx.doi.org/10.1007/JHEP11(2013)215}{{\em JHEP} {\bfseries 11}
  (2013) 215},
\href{http://arxiv.org/abs/1308.6338}{{\ttfamily arXiv:1308.6338 [hep-ph]}}.

\bibitem{Rubio:2014wta}
J.~Rubio and M.~Shaposhnikov, ``{Higgs-Dilaton cosmology: Universality versus
  criticality},'' \href{http://dx.doi.org/10.1103/PhysRevD.90.027307}{{\em
  Phys. Rev.} {\bfseries D90} (2014) 027307},
\href{http://arxiv.org/abs/1406.5182}{{\ttfamily arXiv:1406.5182 [hep-ph]}}.

\bibitem{Boels:2015pta}
R.~H. Boels and W.~Wormsbecher, ``{Spontaneously broken conformal invariance in
  observables},''
\href{http://arxiv.org/abs/1507.08162}{{\ttfamily arXiv:1507.08162 [hep-th]}}.

\bibitem{Karam:2015jta}
A.~Karam and K.~Tamvakis, ``{Dark matter and neutrino masses from a
  scale-invariant multi-Higgs portal},''
  \href{http://dx.doi.org/10.1103/PhysRevD.92.075010}{{\em Phys. Rev.}
  {\bfseries D92} no.~7, (2015) 075010},
\href{http://arxiv.org/abs/1508.03031}{{\ttfamily arXiv:1508.03031 [hep-ph]}}.

\bibitem{DiVecchia:2015jaq}
P.~Di~Vecchia, R.~Marotta, M.~Mojaza, and J.~Nohle, ``{New soft theorems for
  the gravity dilaton and the Nambu-Goldstone dilaton at subsubleading
  order},'' \href{http://dx.doi.org/10.1103/PhysRevD.93.085015}{{\em Phys.
  Rev.} {\bfseries D93} no.~8, (2016) 085015},
\href{http://arxiv.org/abs/1512.03316}{{\ttfamily arXiv:1512.03316 [hep-th]}}.

\bibitem{Karananas:2016grc}
G.~K. Karananas and M.~Shaposhnikov, ``{Scale invariant alternatives to general
  relativity. II. Dilaton properties},''
  \href{http://dx.doi.org/10.1103/PhysRevD.93.084052}{{\em Phys. Rev.}
  {\bfseries D93} no.~8, (2016) 084052},
\href{http://arxiv.org/abs/1603.01274}{{\ttfamily arXiv:1603.01274 [hep-th]}}.

\bibitem{Ferreira:2016vsc}
P.~G. Ferreira, C.~T. Hill, and G.~G. Ross, ``{Scale-Independent Inflation and
  Hierarchy Generation},''
  \href{http://dx.doi.org/10.1016/j.physletb.2016.10.036}{{\em Phys. Lett.}
  {\bfseries B763} (2016) 174--178},
\href{http://arxiv.org/abs/1603.05983}{{\ttfamily arXiv:1603.05983 [hep-th]}}.

\bibitem{Bianchi:2016viy}
M.~Bianchi, A.~L. Guerrieri, Y.-t. Huang, C.-J. Lee, and C.~Wen, ``{Exploring
  soft constraints on effective actions},''
  \href{http://dx.doi.org/10.1007/JHEP10(2016)036}{{\em JHEP} {\bfseries 10}
  (2016) 036},
\href{http://arxiv.org/abs/1605.08697}{{\ttfamily arXiv:1605.08697 [hep-th]}}.

\bibitem{Karananas:2016kyt}
G.~K. Karananas and J.~Rubio, ``{On the geometrical interpretation of
  scale-invariant models of inflation},''
  \href{http://dx.doi.org/10.1016/j.physletb.2016.08.037}{{\em Phys. Lett.}
  {\bfseries B761} (2016) 223--228},
\href{http://arxiv.org/abs/1606.08848}{{\ttfamily arXiv:1606.08848 [hep-ph]}}.

\bibitem{Karam:2016rsz}
A.~Karam and K.~Tamvakis, ``{Dark Matter from a Classically Scale-Invariant
  $SU(3)_X$},'' \href{http://dx.doi.org/10.1103/PhysRevD.94.055004}{{\em Phys.
  Rev.} {\bfseries D94} no.~5, (2016) 055004},
\href{http://arxiv.org/abs/1607.01001}{{\ttfamily arXiv:1607.01001 [hep-ph]}}.

\bibitem{Ferreira:2016wem}
P.~G. Ferreira, C.~T. Hill, and G.~G. Ross, ``{Weyl Current, Scale-Invariant
  Inflation and Planck Scale Generation},''
  \href{http://dx.doi.org/10.1103/PhysRevD.95.043507}{{\em Phys. Rev.}
  {\bfseries D95} no.~4, (2017) 043507},
\href{http://arxiv.org/abs/1610.09243}{{\ttfamily arXiv:1610.09243 [hep-th]}}.

\bibitem{Ferreira:2016kxi}
P.~G. Ferreira, C.~T. Hill, and G.~G. Ross, ``{No fifth force in a scale
  invariant universe},''
  \href{http://dx.doi.org/10.1103/PhysRevD.95.064038}{{\em Phys. Rev.}
  {\bfseries D95} no.~6, (2017) 064038},
\href{http://arxiv.org/abs/1612.03157}{{\ttfamily arXiv:1612.03157 [gr-qc]}}.

\bibitem{Rubio:2017gty}
J.~Rubio and C.~Wetterich, ``{Emergent scale symmetry: Connecting inflation and
  dark energy},'' \href{http://dx.doi.org/10.1103/PhysRevD.96.063509}{{\em
  Phys. Rev.} {\bfseries D96} no.~6, (2017) 063509},
\href{http://arxiv.org/abs/1705.00552}{{\ttfamily arXiv:1705.00552 [gr-qc]}}.

\bibitem{DiVecchia:2017uqn}
P.~Di~Vecchia, R.~Marotta, and M.~Mojaza, ``{Double-soft behavior of the
  dilaton of spontaneously broken conformal invariance},''
\href{http://arxiv.org/abs/1705.06175}{{\ttfamily arXiv:1705.06175 [hep-th]}}.

\bibitem{Guerrieri:2017ujb}
A.~L. Guerrieri, Y.-t. Huang, Z.~Li, and C.~Wen, ``{On the exactness of soft
  theorems},''
\href{http://arxiv.org/abs/1705.10078}{{\ttfamily arXiv:1705.10078 [hep-th]}}.

\bibitem{Tokareva:2017nng}
A.~Tokareva, ``{A minimal scale invariant axion solution to the strong
  CP-problem},''
\href{http://arxiv.org/abs/1705.10836}{{\ttfamily arXiv:1705.10836 [hep-ph]}}.

\bibitem{Gillioz:2017ooc}
M.~Gillioz, ``{Spontaneous conformal symmetry breaking and a massless Wu-Yang
  monopole},''
\href{http://arxiv.org/abs/1707.05325}{{\ttfamily arXiv:1707.05325 [hep-th]}}.

\bibitem{Polchinski:1987dy}
J.~Polchinski, ``{Scale and Conformal Invariance in Quantum Field Theory},''
\href{http://dx.doi.org/10.1016/0550-3213(88)90179-4}{{\em Nucl. Phys.}
  {\bfseries B303} (1988) 226}.

\bibitem{Luty:2012ww}
M.~A. Luty, J.~Polchinski, and R.~Rattazzi, ``{The $a$-theorem and the
  Asymptotics of 4D Quantum Field Theory},''
  \href{http://dx.doi.org/10.1007/JHEP01(2013)152}{{\em JHEP} {\bfseries 01}
  (2013) 152},
\href{http://arxiv.org/abs/1204.5221}{{\ttfamily arXiv:1204.5221 [hep-th]}}.

\bibitem{Fortin:2012hn}
J.-F. Fortin, B.~Grinstein, and A.~Stergiou, ``{Limit Cycles and Conformal
  Invariance},'' \href{http://dx.doi.org/10.1007/JHEP01(2013)184}{{\em JHEP}
  {\bfseries 01} (2013) 184},
\href{http://arxiv.org/abs/1208.3674}{{\ttfamily arXiv:1208.3674 [hep-th]}}.

\bibitem{Nakayama:2013is}
Y.~Nakayama, ``{Scale invariance vs conformal invariance},''
  \href{http://dx.doi.org/10.1016/j.physrep.2014.12.003}{{\em Phys. Rept.}
  {\bfseries 569} (2015) 1--93},
\href{http://arxiv.org/abs/1302.0884}{{\ttfamily arXiv:1302.0884 [hep-th]}}.

\bibitem{Weinberg:1995mt}
S.~Weinberg, {\em {The Quantum theory of fields. Vol. 1: Foundations}}.
\newblock Cambridge University Press,
2005.
\newblock

\bibitem{Weinberg:1996kr}
S.~Weinberg, {\em {The quantum theory of fields. Vol. 2: Modern applications}}.
\newblock Cambridge University Press,
2013.
\newblock

\bibitem{Mack:1976pa}
G.~Mack, ``{Convergence of Operator Product Expansions on the Vacuum in
  Conformal Invariant Quantum Field Theory},''
\href{http://dx.doi.org/10.1007/BF01609130}{{\em Commun. Math. Phys.}
  {\bfseries 53} (1977) 155}.

\bibitem{Pappadopulo:2012jk}
D.~Pappadopulo, S.~Rychkov, J.~Espin, and R.~Rattazzi, ``{OPE Convergence in
  Conformal Field Theory},''
  \href{http://dx.doi.org/10.1103/PhysRevD.86.105043}{{\em Phys. Rev.}
  {\bfseries D86} (2012) 105043},
\href{http://arxiv.org/abs/1208.6449}{{\ttfamily arXiv:1208.6449 [hep-th]}}.

\bibitem{Luscher:1975js}
M.~Luscher, ``{Operator product expansions on the vacuum in conformal quantum
  field theory in two spacetime dimensions},''
\href{http://dx.doi.org/10.1007/BF01608553}{{\em Commun. Math. Phys.}
  {\bfseries 50} (1976) 23--52}.

\bibitem{ElShowk:2011ag}
S.~El-Showk and K.~Papadodimas, ``{Emergent Spacetime and Holographic CFTs},''
  \href{http://dx.doi.org/10.1007/JHEP10(2012)106}{{\em JHEP} {\bfseries 10}
  (2012) 106},
\href{http://arxiv.org/abs/1101.4163}{{\ttfamily arXiv:1101.4163 [hep-th]}}.

\bibitem{Kharzeev:2007wb}
D.~Kharzeev and K.~Tuchin, ``{Bulk viscosity of QCD matter near the critical
  temperature},'' \href{http://dx.doi.org/10.1088/1126-6708/2008/09/093}{{\em
  JHEP} {\bfseries 09} (2008) 093},
\href{http://arxiv.org/abs/0705.4280}{{\ttfamily arXiv:0705.4280 [hep-ph]}}.

\bibitem{Karsch:2007jc}
F.~Karsch, D.~Kharzeev, and K.~Tuchin, ``{Universal properties of bulk
  viscosity near the QCD phase transition},''
  \href{http://dx.doi.org/10.1016/j.physletb.2008.01.080}{{\em Phys. Lett.}
  {\bfseries B663} (2008) 217--221},
\href{http://arxiv.org/abs/0711.0914}{{\ttfamily arXiv:0711.0914 [hep-ph]}}.

\bibitem{Salam:1970qk}
A.~Salam and J.~A. Strathdee, ``{Nonlinear realizations. 2. Conformal
  symmetry},''
\href{http://dx.doi.org/10.1103/PhysRev.184.1760}{{\em Phys. Rev.} {\bfseries
  184} (1969) 1760--1768}.

\bibitem{Volkov:1973vd}
D.~V. Volkov, ``{Phenomenological Lagrangians},''
{\em Fiz. Elem. Chast. Atom. Yadra} {\bfseries 4} (1973) 3--41.

\bibitem{Ogievetsky:1974}
V.~I. Ogievetsky, ``{Nonlinear realizations of internal and space-time
  symmetries},'' {\em in the X-th winter school of theoretical physics in
  Karpacz, Poland} (1974) .

\bibitem{Ivanov:1975zq}
E.~A. Ivanov and V.~I. Ogievetsky, ``{The Inverse Higgs Phenomenon in Nonlinear
  Realizations},''
\href{http://dx.doi.org/10.1007/BF01028947}{{\em Teor. Mat. Fiz.} {\bfseries
  25} (1975) 164--177}.

\bibitem{Low:2001bw}
I.~Low and A.~V. Manohar, ``{Spontaneously broken space-time symmetries and
  Goldstone's theorem},''
  \href{http://dx.doi.org/10.1103/PhysRevLett.88.101602}{{\em Phys. Rev. Lett.}
  {\bfseries 88} (2002) 101602},
\href{http://arxiv.org/abs/hep-th/0110285}{{\ttfamily arXiv:hep-th/0110285
  [hep-th]}}.

\bibitem{Goldstone:1962es}
J.~Goldstone, A.~Salam, and S.~Weinberg, ``{Broken Symmetries},''
\href{http://dx.doi.org/10.1103/PhysRev.127.965}{{\em Phys. Rev.} {\bfseries
  127} (1962) 965--970}.

\bibitem{Higashijima:1994}
K.~Higashijima, ``Nambu-goldstone theorem for conformal symmetry,'' {\em
  Proceedings of XX International Colloquium on Group Theoretical Methods in
  Physics} (1994) 223--228.

\bibitem{Dothan1972}
Y.~Dothan and E.~Gal-Ezer, ``Generalizations of the goldstone and the coleman
  theorems,'' \href{http://dx.doi.org/10.1007/BF02729558}{{\em Il Nuovo Cimento
  A (1971-1996)} {\bfseries 12} no.~2, (Nov, 1972) 465--479}.
  \url{https://doi.org/10.1007/BF02729558}.

\bibitem{Ferrari:1973eq}
R.~Ferrari, ``{On goldstone's theorem for a class of currents not covariant
  under translations},''
\href{http://dx.doi.org/10.1007/BF02728960}{{\em Nuovo Cim.} {\bfseries A14}
  (1973) 386--402}.

\bibitem{Osborn:1993cr}
H.~Osborn and A.~C. Petkou, ``{Implications of conformal invariance in field
  theories for general dimensions},''
  \href{http://dx.doi.org/10.1006/aphy.1994.1045}{{\em Annals Phys.} {\bfseries
  231} (1994) 311--362},
\href{http://arxiv.org/abs/hep-th/9307010}{{\ttfamily arXiv:hep-th/9307010
  [hep-th]}}.

\bibitem{Dorigoni:2009ra}
D.~Dorigoni and V.~S. Rychkov, ``{Scale Invariance + Unitarity => Conformal
  Invariance?},''
\href{http://arxiv.org/abs/0910.1087}{{\ttfamily arXiv:0910.1087 [hep-th]}}.

\bibitem{Gromov:2009tv}
N.~Gromov, V.~Kazakov, and P.~Vieira, ``{Exact Spectrum of Anomalous Dimensions
  of Planar N=4 Supersymmetric Yang-Mills Theory},''
  \href{http://dx.doi.org/10.1103/PhysRevLett.103.131601}{{\em Phys. Rev.
  Lett.} {\bfseries 103} (2009) 131601},
\href{http://arxiv.org/abs/0901.3753}{{\ttfamily arXiv:0901.3753 [hep-th]}}.

\bibitem{Gromov:2009bc}
N.~Gromov, V.~Kazakov, A.~Kozak, and P.~Vieira, ``{Exact Spectrum of Anomalous
  Dimensions of Planar N = 4 Supersymmetric Yang-Mills Theory: TBA and excited
  states},'' \href{http://dx.doi.org/10.1007/s11005-010-0374-8}{{\em Lett.
  Math. Phys.} {\bfseries 91} (2010) 265--287},
\href{http://arxiv.org/abs/0902.4458}{{\ttfamily arXiv:0902.4458 [hep-th]}}.

\bibitem{Gromov:2009zb}
N.~Gromov, V.~Kazakov, and P.~Vieira, ``{Exact Spectrum of Planar ${\cal N}=4$
  Supersymmetric Yang-Mills Theory: Konishi Dimension at Any Coupling},''
  \href{http://dx.doi.org/10.1103/PhysRevLett.104.211601}{{\em Phys. Rev.
  Lett.} {\bfseries 104} (2010) 211601},
\href{http://arxiv.org/abs/0906.4240}{{\ttfamily arXiv:0906.4240 [hep-th]}}.

\bibitem{Shaposhnikov:2007nj}
M.~Shaposhnikov, ``{Is there a new physics between electroweak and Planck
  scales?},'' in {\em {Astroparticle Physics: Current Issues, 2007 (APCI07)
  Budapest, Hungary, June 21-23, 2007}}.
\newblock \href{http://arxiv.org/abs/0708.3550}{{\ttfamily arXiv:0708.3550
  [hep-th]}}.
\newblock
\url{http://inspirehep.net/record/759157/files/arXiv:0708.3550.pdf}.
\newblock

\bibitem{Giudice:2013nak}
G.~F. Giudice, ``{Naturalness after LHC8},'' {\em PoS} {\bfseries EPS-HEP2013}
  (2013) 163,
\href{http://arxiv.org/abs/1307.7879}{{\ttfamily arXiv:1307.7879 [hep-ph]}}.

\bibitem{Karananas:2017mxm}
G.~K. Karananas and M.~Shaposhnikov, ``{Gauge coupling unification without
  leptoquarks},'' \href{http://dx.doi.org/10.1016/j.physletb.2017.05.065}{{\em
  Phys. Lett.} {\bfseries B771} (2017) 332--338},
\href{http://arxiv.org/abs/1703.02964}{{\ttfamily arXiv:1703.02964 [hep-ph]}}.

\end{thebibliography}\endgroup
\end{document}